# Faraday rotation echo spectroscopy and detection of quantum fluctuations

Shao-Wen Chen & Ren-Bao Liu*

Department of Physics, Centre for Quantum Coherence, & Institute of Theoretical Physics, The Chinese University of Hong Kong, Hong Kong, China

**Abstract:**

**Central spin decoherence is useful for detecting many-body physics in environments and moreover, the spin echo control can remove the effects of static thermal fluctuations so that the quantum fluctuations are revealed. The central spin decoherence approach, however, is feasible only in some special configurations and often requires uniform coupling between the central spin and individual spins in the baths, which are very challenging in experiments. Here, by making analogue between central spin decoherence and depolarization of photons, we propose a scheme of Faraday rotation echo spectroscopy (FRES) for studying quantum fluctuations in interacting spin systems. The echo control of the photon polarization is realized by flipping the polarization with a birefringence crystal. The FRES, similar to spin echo in magnetic resonance spectroscopy, can suppress the effects of the static magnetic fluctuations and therefore reveal dynamical magnetic fluctuations. We apply the scheme to a rare-earth compound $LiHoF_4$ and calculate the echo signal, which is related to the quantum fluctuations of the system. We observe enhanced signals at the phase boundary. The FRES should be useful for studying quantum fluctuations in a broad range of spin systems, including cold atoms, quantum dots, solid-state impurities, and transparent magnetic materials.**

**Full text:**

When a central spin is coupled to a spin bath, the quantum coherence of the central spin would be lost due to the noise from the bath[1,2,3]. On the one hand, the central spin decoherence is an important issue in quantum coherence based technologies such as quantum computing[4,5] and magnetometry[6-9]. On the other hand, since the noise is caused either by thermal fluctuations or by elementary excitations in the bath, the central spin decoherence is a useful probe of many-body physics in the bath. A number of interesting effects has been discovered by viewing the central spin decoherence as a probe of many-body physics [10-12].

Coupling a single spin to a spin bath, however, is not a trivial task in experiments and is feasible only in certain specially designed systems. Moreover, such coupling between the central spin and individual bath spins is often required to be uniform [10-12], which put additional constraints on experiments. A potential solution is motivated by the spin noise spectroscopy[13-16], where the polarization of photons plays the role of a central spin and the depolarization of the photons due to the spin noise resembles the central spin decoherence. The coupling between photons and spins has been well established for many systems including atoms[17], quantum dots[18,19], and transparent magnetic materials[20]. A famous example is the Faraday rotation [17-21]. Due to slow spatial variation of laser pulses, the coupling can be easily made almost uniform for all spins interacting with the photons. Therefore, in this paper, we propose a scheme based on Faraday rotation and photon depolarization to study many-body physics of spins systems, in lieu of the central spin decoherence method[10-12].

A particularly useful technique in central spin decoherence method, which is not yet available in the optical spin noise spectroscopy, is the spin echo[22], or more generally, dynamical decoupling control[23,24]. In spin echo or dynamical decoupling, the central spin is flipped and effectively senses the noises from opposite directions before and after the flip control. Therefore, if the noise is static (or slow enough), the effect would be cancelled when the time the central spin evolves after the flip equals that before the flip and the central spin coherence will recover at that particular time, resembling an echo. Since the static noises usually result from the classical configurations of the laboratories and from the thermal fluctuations in the baths, the central spin decoherence under the spin echo control is particularly sensitive to the dynamical fluctuations in the bath, which have quantum origins. Therefore, it has been shown before that even at a high temperature, the central spin decoherence, when it is under echo control, can still single out the quantum fluctuations[12]. To incorporate the echo control in the optical noise spectroscopy, we design an optical set up in which the photon polarization can be flipped by a birefringence crystal and therefore the analogue to spin echo can be realized in the photon depolarization measurement. We test our scheme on a rare-earth compound $LiHoF_4$ which is an experimental realization of Ising magnets[25,26] and has sizable Faraday rotation effect on optical probes[20].

## RESULTS

**Fluctuations of Faraday rotation angle**

A central spin-1/2 ($|\uparrow\rangle+|\downarrow\rangle$) under an external field $b(t)$ along the $z$- axis accumulates a phase shift such as ($e^{i\theta}|\uparrow\rangle+|\downarrow\rangle$), with $\theta = \int_0^t b(t)dt$. If the external field is random (due to interaction with fluctuating spins in the bath), the coherence of the spin will be lost.

In analogue to this spin decoherence, we consider depolarization of a linearly polarized laser pulse. Being a superposition of two circular polarizations ($\sigma^+ + \sigma^-$), the laser pulse can play the role of a central spin. When the linearly polarized laser pulse propagates through a magnetic sample that contains fluctuating spins ($\mathbf{J}_i$) [see Figure 1 (a)], the two circular components get different phase shifts ($e^{i\theta}\sigma^+ + e^{-i\theta}\sigma^-$), which produce a rotation angle $\theta$ of the linear polarization, known as the Faraday rotation (FR). The FR angle is proportional to the magnetization along the direction of propagation and the thickness ($L_S$) of the sample. Here we consider the case that the pulse length is much longer than the thickness of the transparent sample. The laser, therefore, can be considered to be interacting with all the spins within the cross sectional area ($A$) simultaneously. The FR angle gained by a linearly polarized laser after it travels through the sample along the $z$ direction is

$$\theta(t_p) = \alpha t_p^{-1} L_S \int_0^{t_p} M_z(t) dt, \tag{1}$$

where $t_p$ is the duration of the laser pulse, $M_z(t) = N^{-1}\sum_{i=1}^{N} J_i^z(t)$ is the magnetization, $N = \rho A L_S$ is the number of spins located within the interaction region ($\rho$ being the density of spins), and $\alpha$ is a coupling coefficient, which depends on the laser frequency and the material parameters (see Methods).

The fluctuation of the magnetization causes a random phase shift of the circular polarized components of the laser, leading to the depolarization of the laser, which is similar to the decoherence of a spin-1/2 under a random field. The fluctuation of the FR angle is determined by the fluctuation of the magnetization $C_z(t_1 - t_2) \equiv \langle \Delta M_z(t_1) \Delta M_z(t_2) \rangle$ as (see Methods)

$$\langle \Delta \theta_{\text{FR}}^2(t_p) \rangle = \alpha^2 t_p^{-2} L_S^2 \iint_0^{t_p} dt_1 dt_2 C_z(t_1 - t_2), \tag{2}$$

where $\Delta \theta_{\text{FR}} \equiv \theta_{\text{FR}} - \langle \theta_{\text{FR}} \rangle$ and $\Delta M_z \equiv M_z - \langle M_z \rangle$. The symbol $\langle ... \rangle$ denotes the ensemble average, that is, $\langle M \rangle = \sum_n P_n \langle \psi_n | M | \psi_n \rangle$, where $|\psi_n\rangle$ and $E_n$ are, respectively, the eigenstate and eigenenergy of the

spin system, and $P_n = e^{-\beta E_n} / \sum_m e^{-\beta E_m}$ is the probability distribution, with $\beta$ being the inverse temperature. The Fourier transform of the correlation function $C_z(t)$ gives the noise spectrum of the spin system $S(\omega) \equiv \int C_z(t) \exp(i\omega t) dt$.

The noise spectrum can be understood as caused by two mechanisms. One part of the fluctuations is caused by the transitions between different energy eigenstates. The corresponding spectrum is $S_Q(\omega) = 2\pi \sum_{n \neq m} \delta[\omega - (E_n - E_m)] P_n |\langle \psi_n | M | \psi_m \rangle|^2$, which contains non-zero frequency components. Therefore, it is dynamical and quantum. Another part of the fluctuations is originated from the probability distribution $P_n$ at finite temperature and has only the zero frequency component:

$S_{th}(\omega) = 2\pi \delta(\omega) \left( \sum_n P_n \langle \psi_n | M | \psi_n \rangle^2 - \langle M \rangle^2 \right)$, which is the static thermal fluctuation and vanishes at zero temperature. The effects of the static thermal fluctuations can be removed by spin echo[22], or by the Faraday rotation echo as studied in this paper. Therefore the features of the quantum fluctuations would be revealed by the echo methods.

**Faraday rotation echo spectroscopy (FRES)**

The idea of FRES is explained as follows. First we let the laser pulse go through the sample and accumulate an FR angle ($\theta$) by interacting with the spins in the sample. After the laser pulse interacting with the sample, it is reflected and passes through a linear birefringence crystal[21]. The birefringence crystal has different refractive indices ($n_o$ and $n_e$) for the linearly polarized light with polarization parallel and perpendicular to the optical axis of the crystal. The original laser polarization (before interacting with the sample) is set to be parallel or perpendicular to the optical axis of the crystal. If the thickness of the crystal ($L_{bc}$) is chosen such that the two orthogonal polarizations accumulate a phase difference $\pi$, namely,

$$2\pi(n_o - n_e)\frac{2L_{bc}}{\lambda} = \pi, \tag{3}$$

where $\lambda$ is the wavelength of the laser in the vacuum, the polarization of the laser would be flipped after propagating through the birefringence crystal and the Faraday rotation accumulated before would change its sign ($\theta \to -\theta$) [see Figure 1 (c-d)]. This is a straightforward analogue to the spin flip in spin echo. Then, we let the laser pulse go through the sample again and interact with the spins once more [see Figure 1 (c)]. As compared to the spin echo, this Faraday rotation echo can have an additional time delay $\tau$ between the interactions with the sample before and after interaction with the birefringence crystal. By defining the modulation function [Figure 1 (d)]

$$f_{echo}(t,\tau) = \begin{cases} -1 & \text{for } t \in [0, t_p], \\ +1 & \text{for } t \in [t_p + \tau, 2t_p + \tau], \\ 0 & \text{else,} \end{cases} \tag{4}$$

the FR angle of the final output laser reads

$$\theta_{echo}(2t_p) = \alpha t_p^{-1} L_S \int_0^{2t_p+\tau} M_z(t) f_{echo}(t,\tau) dt. \tag{5}$$

If the magnetization fluctuation of the spin system is static, the net FR angle after the echo would be zero. Therefore, the final FR measures the dynamical fluctuation of the magnetization, which is related to quantum transitions in the spin system.

We define $f_{\text{FID}}(t,\tau)=1$ for $t\in[0,t_p]$ [see Figure 1 (b)] as the modulation function for the case of no echo control (similar to the free-induction decay in magnetic resonance spectroscopy). The expressions of the FR angle in the two different cases (echo and free-induction decay) are unified as

$$\theta_\eta(T_\eta) = \alpha t_p^{-1} L_S \int_0^{T_\eta+\tau} M_z(t) f_\eta(t,\tau) dt, \qquad (6)$$

where $\eta=$ "FID", or "echo" indicates the free-induction decay or echo configurations, respectively. $T_\eta$ is the total time that the laser interacts with the spin system (i.e., $T_{\text{FID}} = t_p$ and $T_{\text{echo}} = 2t_p$). The fluctuation of the FR angle is determined by the magnetic noise spectrum as[27]

$$\begin{aligned}\langle \Delta\theta_\eta^2(T_\eta) \rangle &= \alpha^2 t_p^{-2} L_S^2 \iint_0^{T_\eta+\tau} dt_1 dt_2 C_z(t_1-t_2) f_\eta(t_1,\tau) f_\eta(t_2,\tau) \\ &= \alpha^2 \frac{L_S^2}{t_p^2} \iint_0^{T_\eta+\tau} C_z(0) f_\eta(t_1,\tau) f_\eta(t_2,\tau) dt_1 dt_2 + \alpha^2 \frac{L_S^2}{t_p^2} \int \frac{d\omega}{2\pi} S_Q(\omega) \frac{|F_\eta(\omega)|^2}{\omega^2},\end{aligned} \qquad (7)$$

where the filter function $\omega^{-2}|F_\eta(\omega)|^2$ is the Fourier transform of the modulation function, with

$$F_\eta(\omega) \equiv \omega \int f_\eta(t,\tau) e^{i\omega t} dt . \qquad (8)$$

The expression of $\langle \Delta\theta_{\text{echo}}^2 \rangle$ in equation (7) is similar to the phase fluctuation of spins in magnetic resonance spectroscopy. The first term in equation (7) comes from the thermal fluctuation [$S_{\text{th}}(\omega)$], which vanishes in the case of FR echo since $C_z(0)$ is a constant. Therefore, the FR echo measures the quantum fluctuation of the magnetization. The tunable delay time $\tau$ adds extra flexibility for studying the quantum fluctuations, as compared with the conventional spin echo. For example, a large delay time can be used to single out the effect of low energy excitations.

The fluctuation of the FR angle will result in depolarization of the laser pulse. The degree of polarization of the laser pulse after the interaction with the sample is

$$P \equiv \frac{I_{max} - I_{min}}{I_{max} + I_{min}} = 1 - 2\langle \Delta\theta_\eta^2 \rangle, \quad (9)$$

where $I_{max} \propto \langle \cos^2 \Delta\theta_\eta \rangle$ and $I_{min} \propto \langle \sin^2 \Delta\theta_\eta \rangle$ are the maximum and minimum intensities of the light passes through a linear polarizer, respectively. Therefore, the depolarization is directly related to the fluctuations of the FR angle.

## FRES of LiHoF$_4$

The lithium rare-earth tetrafluorides are a family of compounds used as a testing ground for the physics of spin models. All of these compounds are optically transparent, which makes them ideal for optical studies[20]. We choose the lithium holmium tetrafluoride crystal LiHoF$_4$[26,28,29] as our model system in this study.

LiHoF$_4$ has a scheelite structure [Figure 2 (a)] with the lattice constants[30] $a = 5.175$ Å and $c = 10.75$ Å. The magnetic properties of this compound come from the Holmium ions (Ho$^{+3}$), which can be treated as a system of spin-8. The interaction of the Ho$^{+3}$ ions with the surrounding Li$^+$ and F$^-$ ions are described by a crystal-field Hamiltonian $H_{CF}$ (see Methods). The crystal field produces an energy level splitting of the Holmium ions. The Holmium spin has a ground state doublet and 15 excited states. The lowest excited energy level lies about 11 Kelvin above the ground state. When the temperature is much lower than 11 Kelvin, only the ground state doublets are non-negligibly populated. According to the expression of the Steven operators (see Methods), the dominant term in the crystal field Hamiltonian is $\sim -(J_i^z)^2$ (the z-direction is along the **c** axis of the crystal), which means that the ground state doublet are basically $|8,-8\rangle$ and $|8,+8\rangle$. The flipping between these two states needs to go through the excited states. Therefore, at temperature much lower than 11 Kelvin, the spins effectively have the Ising type of interaction. Quantum fluctuations of the spins can be induced by applying a transverse magnetic field $B_x$ along the x-axis.

The full Hamiltonian of LiHoF$_4$ is[26,28,29]

$$H = \sum_i [H_{CF}(\mathbf{J}_i) + \mathcal{A}\mathbf{J}_i \mathbf{I}_i - g\mu_B \mathbf{J}_i \mathbf{B}] - \frac{1}{2}\sum_{ij} \mathcal{J}_D \mathbf{J}_i \ddot{\mathcal{D}}_{ij} \mathbf{J}_j - \frac{1}{2}\sum_{<ij>} \mathcal{J}_{12} \mathbf{J}_i \mathbf{J}_j, \qquad (10)$$

where $\mathbf{J}_i$ and $\mathbf{I}_i$ are the electron spin ($J=8$ with $g=5/4$) and the nuclear spin ($I=7/2$) of the $i$-th ion, respectively. The magnetic interactions in LiHoF$_4$ include the long-range dipole interaction between the Holmium magnetic moments, with $[\ddot{\mathcal{D}}_{ij}]_{\alpha,\beta} = D_{\alpha\beta}(ij)$ being the dipole sum, the exchange interaction ($\mathcal{J}_{12} = 0.6\ \mu eV \sim 0.007$ Kelvin) between Holmium spins in the nearest neighbors, and the isotropic hyperfine interaction between the nuclear and electron magnetic moments on the same site ($\mathcal{A} = 3.36\ \mu eV \sim 0.04$ Kelvin). In the case of zero field, LiHoF$_4$ forms a ferromagnetic order at the critical temperature 1.53 Kelvin. When a transverse field $B_x$ is applied, the ordering temperature gradually approaches to zero at the critical field $B_C = 5$ Tesla. In Figure 2 (b), the expectation value of the single site angular momentum operator $J_i^z$ is calculated with the mean-field approximation using the full Hamiltonian in equation (10). The appearance of the spontaneous magnetization signatures the emergence of the ferromagnetic order. The paramagnetic-ferromagnetic phase boundary agrees with the phase diagram obtained in previous works (see, e.g., Ref. 31). Figure 2 (c-e) shows the mean field results of the correlation function of the magnetic fluctuations $\langle \Delta J_i^z(t) \Delta J_i^z \rangle$ as a function of the transverse field $B_x$ and the time $t$ for various temperatures. In the long time limit (in which the time is longer than the inverse energy gap, here $t > 0.1$ ns in our estimation)[32], the onset of oscillations in the paramagnetic phase indicates the phase boundary in Figure 2 (b).

The FR echo signal is evaluated according to equation (7). The correlation function $C_z$ is obtained with the mean-field approximation using the full Hamiltonian, which is related to the single spin correlation functions by

$$C_z(t) = \langle \Delta M_z(0) \Delta M_z(t) \rangle = N^{-1} \langle \Delta J_i^z(0) \Delta J_i^z(t) \rangle. \tag{11}$$

Note that in the mean-field approximation, the correlations between spins at different sites vanish (i.e., $\langle \Delta J_i^z(0) \Delta J_j^z(t) \rangle = 0$ for $i \neq j$). Figure 3 (a) plots the FR angle fluctuation under the free-induction decay configuration as a function of the temperature and the transverse field. When the system is in the paramagnetic phase, the fluctuation of the magnetization is suppressed by the transverse field and is less sensitive to $B_x$. A critical feature, namely, the sudden change of slope [see inset of Figure 3 (a)], appears at the phase boundary. Figure 3(b) shows the FR echo signal (for delay time $\tau = 0$). After the thermal fluctuation is removed by the echo control, a peak feature is observed at the phase boundary, where the quantum fluctuation diverges [see inset of Fig 3 (b)]. According to the estimation (see Methods), for a laser with wavelength $\lambda = 435$ nm and cross-section area $A = 10$ μm$^2$ the strength of the echo signal is about $10^{-9}$ (degree)$^2$, which is experimentally observable (see e.g., Ref. 33). It should be pointed out that the mean-field approximation used in the calculation underestimates the fluctuation near the phase boundary since the long-range correlations of the fluctuations emerge at the phase transitions but the mean-field approximation considers only the local correlations.

# DISCUSSION

To better understand the relation between the phase transitions and the features in the FRES, we study the magnetic noise spectra of the LiHoF$_4$ system. Figure 4 (a) shows some typical examples of the noise spectra at temperature 0.9 Kelvin. The gap of the system reaches the minimal value ($\sim 6.7$ $\mu$eV) around $B_x = 3.2$ Tesla [see Figure 4 (b)], which indicates the onset of the phase transition. Therefore, the low frequency components of $S_Q$ have larger contribution to the Faraday rotation fluctuation when the external field is closer to the critical point. Since we are considering the case of finite temperature, there also exist some lower frequency components of noise spectra due to the transitions between the excited states. But the contributions from those components are very small due to the small distribution probabilities of the excited states. Therefore, the critical features around the phase transition point are mainly determined by the transition between the ground state and the lowest excited state [as indicated by the purple dashed circle in Figure 4 (a-b)], when the interacting time is comparable to or longer than the inverse excitation gap (the "long-time" limit).

Figure 4 (c) shows the dependence of the peak features of $\langle \Delta \theta_{\text{echo}}^2 \rangle$ on the total interaction time $T_{\text{echo}}$ between the laser pulse and sample. When $T_{\text{echo}}$ is long enough, a peak feature emerges at the phase transition point. The peak becomes sharper as the interacting time $T_{\text{echo}}$ further increases. The interaction time $T_\eta = 0.3$ ns in Figure 3 (a-b) is chosen long enough for the critical features to pronounce.

As compared with the conventional spin echo in magnetic resonance spectroscopy, the FRES has extra controllability to engineer the filter function by tuning the delay time between the interaction intervals. To show this controllability, we study the effect of the delay time. Figure 5 (a) plots the FR echo signal with a delay of 0.25 ns. The effect of the delay time could be considered as expanding the modulation function to a longer time interval and making the low frequency excitations more important

[see Figure 5 (b)]. Therefore, the critical feature at the phase boundary persists as we increase the delay time [see Figure 5 (c)]. In addition, a long delay time can also reveal the critical features around the phase transition points. In Figure 5 (d), the FR echo signal of $T_{echo} = 0.1$ ns is plotted with different delay times. By prolonging the delay time, a sharp peak emerges at the phase transition point.

The FRES scheme can be straightforwardly extended to more complicated configurations, corresponding to different kinds of dynamical decoupling sequences[23,24], by letting the laser pulse interact with the sample and the birefringence crystal multiple times.

In this paper we have assumed a square shape for the laser pulse. In general, by shaping the laser pulse one can realize more complicated modulation functions (which would be particularly useful for, e.g., spectroscopy of the spin noises in the bath[34,35]).

# METHODS

## Faraday Rotation

For simplicity, we consider the FR of a linearly polarized laser weakly coupled to a single spin **J**. The Faraday rotation is originated from the relative phase shift of the two circular polarized mode (with photon annihilation operators $b_+$ and $b_-$). The effect Hamiltonian is

$$V = \chi J^z (b_+^\dagger b_+ - b_-^\dagger b_-), \quad (12)$$

where $\chi \propto (At_p)^{-1}$ is the coupling strength. The weak coupling condition is $\chi t_p \ll 1$. The initial state of the whole system is on a factorized state

$$|\phi_0\rangle = e^{Cb_H^\dagger - C^* b_H} |0\rangle \otimes |J, m\rangle, \quad (13)$$

where $b_H = (b_+ + b_-)/\sqrt{2}$ is the annihilation operator of the linear horizontal mode, that is, the laser is initially on a linearly polarized coherent state, and $|J, m\rangle$ is an eigenstate of $J^z$, with eigenvalue $m$. The evolution of the state is

$$|\phi(t)\rangle = e^{Cb^\dagger_{\theta_m(t)} - \text{H.c.}} |0\rangle \otimes |J, m\rangle, \quad (14)$$

where $b_{\theta_m} \equiv \cos\theta_m b_H - \sin\theta_m b_V$ is the annihilation operator of the linearly polarized mode with the polarization rotated by an angle $\theta_m(t) = \chi m t$. Here, $b_V = \frac{1}{\sqrt{2}i}(b_+ - b_-)$ is the annihilation operator of the vertically polarized mode. The ensemble measurement of the FR angle gives

$$\langle \theta(t) \rangle = \chi \langle J_z \rangle t. \quad (15)$$

When a linearly polarized laser couples to $N$ fluctuating spins with Hamiltonian

$$V(t) = \chi N M_z(t)(b_+^\dagger b_+ - b_-^\dagger b_-), \quad (16)$$

the Faraday rotation angle is

$$\theta(t) = \chi N \int_0^t M_z(t') dt', \quad (17)$$

where $M_z(t) = N^{-1} \sum_i J_i^z(t)$ is the magnetization. Considering $\chi \propto (A t_p)^{-1}$ and $N \propto A L_S$, we obtain equation (1).

## Crystal-field Hamiltonian

The crystal-field Hamiltonian is[36]

$$H_{CF} = \sum_{l=2,4,6} B_l^0 O_l^0 + \sum_{l=4,6} B_l^4(c) O_l^4(c) + B_6^4(s) O_6^4(s), \qquad (18)$$

with the Steven's operators $O_l^m$ and the corresponding crystal-field parameters $B_l^m$. The $S_4$ symmetry of the Ho-ion surroundings is responsible for the terms in equation (18). The specific forms of the Steven's operators are listed below.

$$O_2^0 = 3(J_i^z)^2 - J(J+1),$$

$$O_4^0 = 35(J_i^z)^4 - 30J(J+1)(J_i^z)^2 + 25(J_i^z)^2 - 6J(J+1) + 3J^2(J+1)^2,$$

$$O_6^0 = 231(J_i^z)^6 - 315J(J+1)(J_i^z)^4 + 735(J_i^z)^4 + 105J^2(J+1)^2(J_i^z)^2$$
$$- 525J(J+1)(J_i^z)^2 + 294(J_i^z)^2 - 5J^3(J+1)^3 + 40J^2(J+1)^2 - 60J(J+1),$$

$$O_4^4(c) = \frac{1}{2}\left[(J_i^+)^4 + (J_i^-)^4\right],$$

$$O_6^4(c) = \frac{1}{4}\left[(J_i^+)^4 + (J_i^-)^4\right]\left[11(J_i^z)^2 - J(J+1) - 38\right] + \text{H.c.},$$

$$O_6^4(s) = \frac{1}{4i}\left[(J_i^+)^4 - (J_i^-)^4\right]\left[11(J_i^z)^2 - J(J+1) - 38\right] + \text{H.c.},$$

where $J_i^\pm = J_i^x \pm i J_i^y$.

In this paper, we use the same crystal-field parameters as in Ref. 28, as listed below.

$B_2^0 = -60\ \mu\text{eV},\ B_4^0 = 0.35\ \mu\text{eV},\ B_4^4 = 3.6\ \mu\text{eV},\ B_6^0 = 0.0004\ \mu\text{eV},\ B_6^4(c) = 0.07\ \mu\text{eV},\ B_6^4(s) = \pm 0.0098\ \mu\text{eV}.$

**Mean field calculation**

The results we present in Figs. 2-5 are calculated with the mean field theory. The full Hamiltonian of the LiHoF$_4$ system in equation (10) includes two parts. The non-interaction part

$$H_0 = \sum_i [H_{CF}(\mathbf{J}_i) + A\mathbf{J}_i \cdot \mathbf{I}_i - g\mu_B \mathbf{J}_i \cdot \mathbf{B}], \tag{19}$$

which is a single-ion Hamiltonian, and the interaction part

$$H_1 = -\frac{1}{2}\sum_{ij} \mathcal{J}_D \mathbf{J}_i \ddot{\mathcal{D}}_{ij} \mathbf{J}_j - \frac{1}{2}\sum_{<ij>} \mathcal{J}_{12} \mathbf{J}_i \mathbf{J}_j, \tag{20}$$

which is treated with standard mean-field approach (see, e.g., Ref. 36). The mean-field Hamiltonian is obtained by neglecting the two-site fluctuation terms

$$\begin{aligned} H_1^{MF} \approx &-\frac{1}{2}\sum_{ij} \mathcal{J}_D \left[ \langle \mathbf{J}_j \rangle (\mathbf{J}_i \cdot \ddot{\mathcal{D}}_{ij} + \ddot{\mathcal{D}}_{ji} \cdot \mathbf{J}_i) - \langle \mathbf{J}_i \rangle \cdot \langle \mathbf{J}_j \rangle \cdot \ddot{\mathcal{D}}_{ij} \right] \\ &-\frac{1}{2}\sum_{<ij>} \mathcal{J}_{12} \left[ 2\langle \mathbf{J}_j \rangle \cdot \mathbf{J}_i - \langle \mathbf{J}_i \rangle \cdot \langle \mathbf{J}_j \rangle \right] \end{aligned}. \tag{21}$$

After replacing $H_1$ with $H_1^{MF}$, the Hamiltonian of the LiHoF$_4$ system becomes a single-ion Hamiltonian, which is then exactly solved by numerical diagonalization. The mean field $\langle \mathbf{J}_j \rangle$ is obtained by iteration.

### Estimation of the signal strength

The coupling strength $\alpha$ is evaluated with

$$\alpha \approx \frac{\langle \theta \rangle}{\langle M_z \rangle L_S}, \tag{22}$$

for static magnetization. According to Ref. 20, $\theta / L_S \approx 9260$ degree/cm in the case of saturated polarization ($\langle M_z \rangle = 8$) for a laser with wavelength 435 nm. Therefore $\alpha = 1157.5$ degree/cm is used for signal strength estimation in this paper.

Since the correlations between different electron spins are neglected in the mean-field approximation, we have $C_z(t) = N^{-1}\langle \Delta J_i^z(0) J_i^z(t)\rangle$. Therefore,

$$\langle \Delta\theta_\eta^2(T_\eta)\rangle = \frac{\alpha^2 L_S}{\rho A t_p^2} \int\int_0^{T_\eta+\tau} \langle \Delta J^z(t_1)\Delta J^z(t_2)\rangle f_\eta(t_1,\tau) f_\eta(t_2,\tau) dt_1 dt_2, \qquad (23)$$

where $\rho \approx 1.39\times 10^{28} m^{-3}$ is the density of spins in LiHoF$_4$ system. The results in Figures 3-5 are estimated according to the above equation. It should be noted that, the mean-field approximation is only valid away from the critical points. In this paper, the mean-field results are used to illustrate the qualitative properties of the FR echo signals around the phase boundary.

**Acknowledgement:** This work was supported by Hong Kong Research Grants Council General Research Fund Project 401011 and The Chinese University of Hong Kong Focused Investments Scheme.

**Author Contributions** R.B.L. conceived the idea, designed the models and formulated the theories. S.W.C. studied the LiHoF$_4$ system. Both authors analyzed the results and wrote the paper.

**Competing financial interests** The authors declare no competing financial interests. Correspondence and requests for materials should be addressed to R.B.L. (rbliu@phy.cuhk.edu.hk).

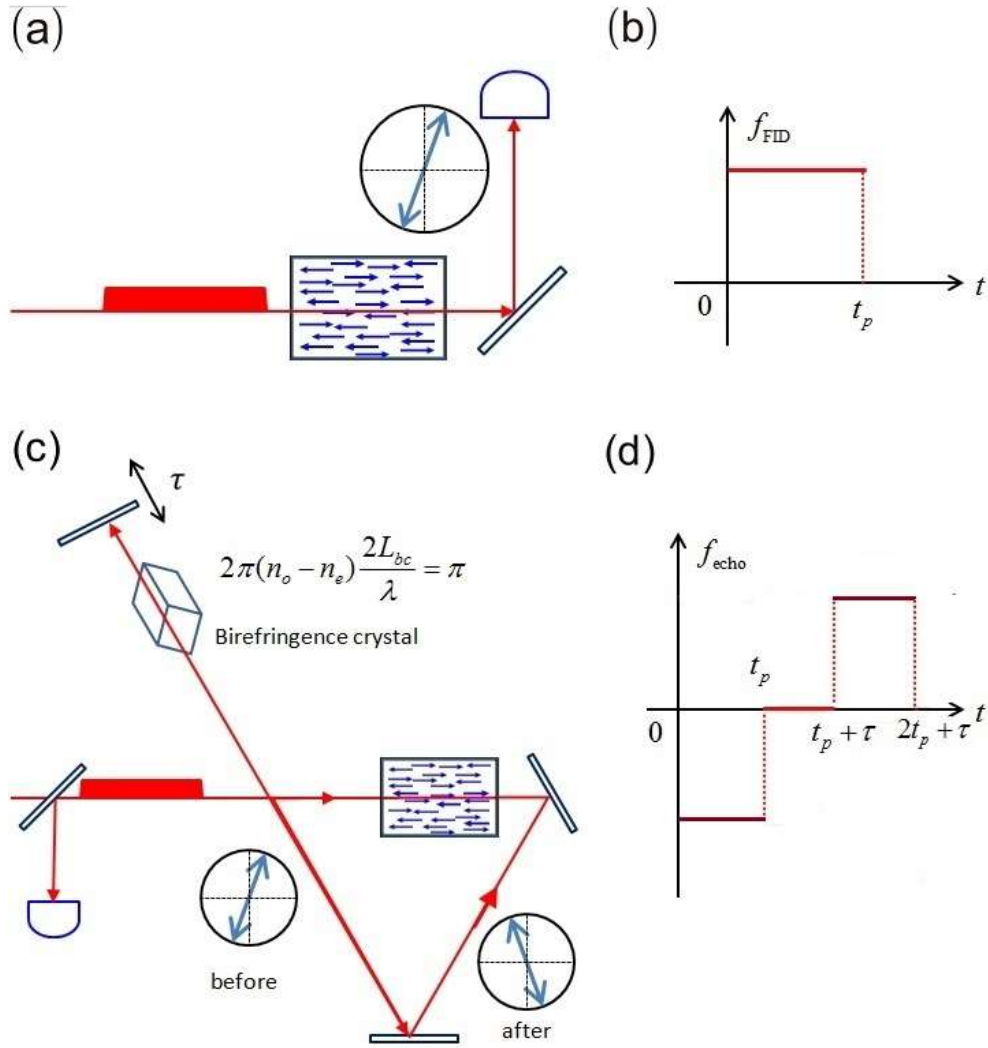

Figure 1 **Schematic scheme of Faraday rotation echo.** (a) A linearly polarized laser pulse passes through a sample containing fluctuating spins and then the FR angle is measured without echo control (the free-induction decay case). (b) Modulation function $f_{FID}$ in the free-induction decay case. (c) The Faraday rotation echo scheme. A linearly polarized laser pulse passes through a sample containing fluctuating spins. After interacting with the sample, the laser pulse passes through a linear birefringence crystal and then is reflected. The birefringence crystal flips the Faraday rotation angle of the laser. Then the laser pulse interacts with the sample again before its polarization is measured. (d) Modulation function $f_{echo}$ in the case of echo control.

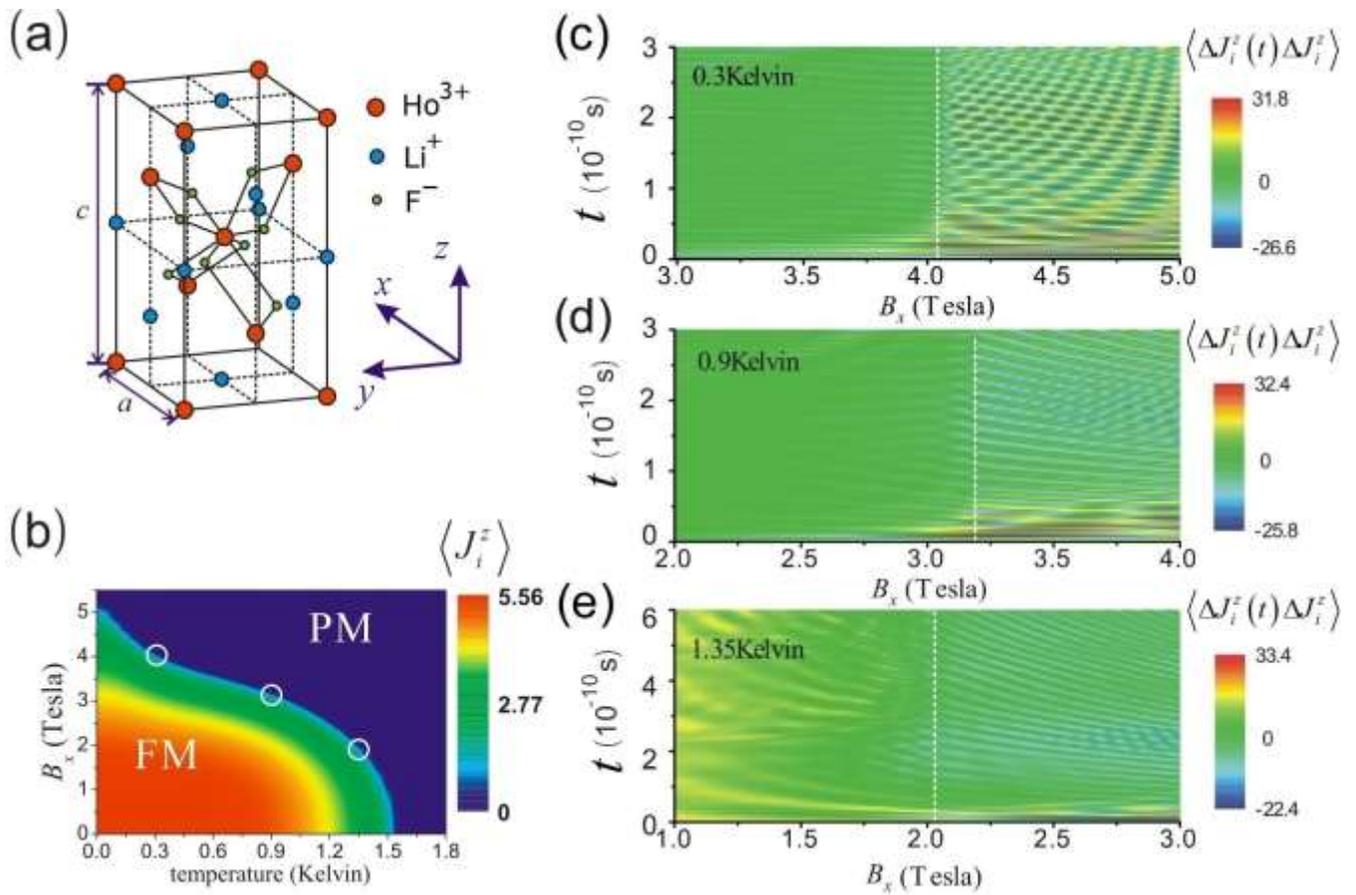

Figure 2 **Phase diagram of LiHoF$_4$ and magnetic correlation functions.** (a) Lattice structure of LiHoF$_4$. (b) Magnetization of LiHoF$_4$ as function of temperature and transverse magnetic field. Sudden appearance of magnetization along the $z$-axis indicates the formation of the ferromagnetic order. The correlation function $\langle \Delta J^z(t) \Delta J^z \rangle$ is plotted as a function of the transverse field $B_x$ and time $t$ at temperature (c) 0.3 Kelvin, (d) 0.9 Kelvin and (d) 1.35 Kelvin. The corresponding critical fields are indicated as open circles in (b) and dashed lines in (c)-(e). An obvious oscillation feature appears in the paramagnetic phase.

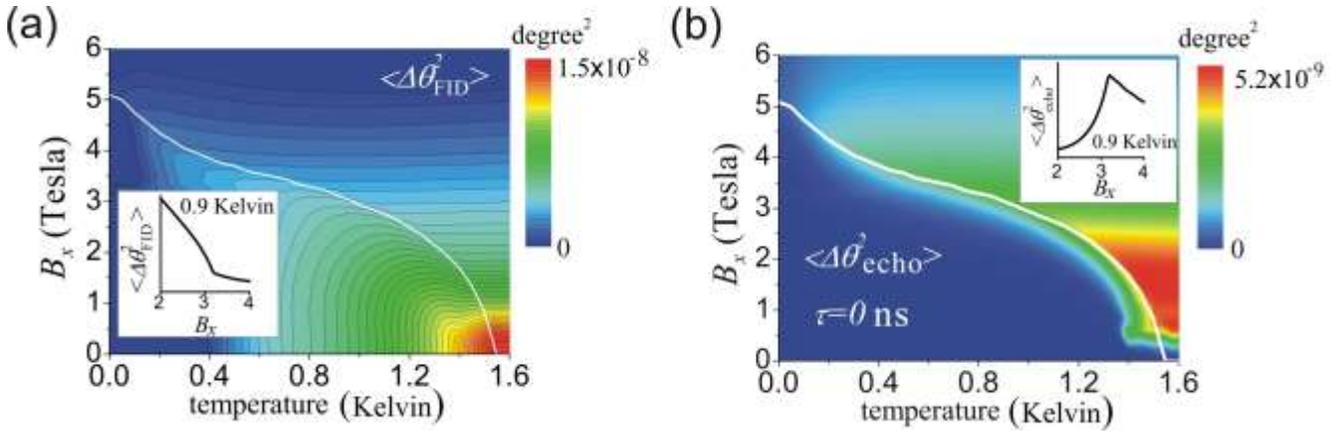

Figure 3 **Fluctuations of the Faraday rotation angle.** (a) The FR angle fluctuation $\langle\Delta\theta_{FID}^2\rangle$ in the free-induction decay configuration (no echo control) is plotted as a function of temperature and transverse field, for interaction time $T_{FID} = 0.3$ ns. The inset shows $\langle\Delta\theta_{FID}^2\rangle$ as a function of the transverse field at temperature 0.9 Kelvin. (b) The fluctuation of the FR angle under echo control $\langle\Delta\theta_{echo}^2\rangle$ as a function of temperature and transverse field, for interaction time $T_{echo} = 0.3$ ns and delay between two interaction intervals $\tau = 0$. The inset shows $\langle\Delta\theta_{echo}^2\rangle$ as a function of the transverse field at temperature 0.9 Kelvin. The sample thickness $L_S = 0.5$ cm, and laser cross sectional area $A = 10$ μm$^2$ are assumed in all the estimations. The white curves in (a) and (b) indicate the phase boundary.

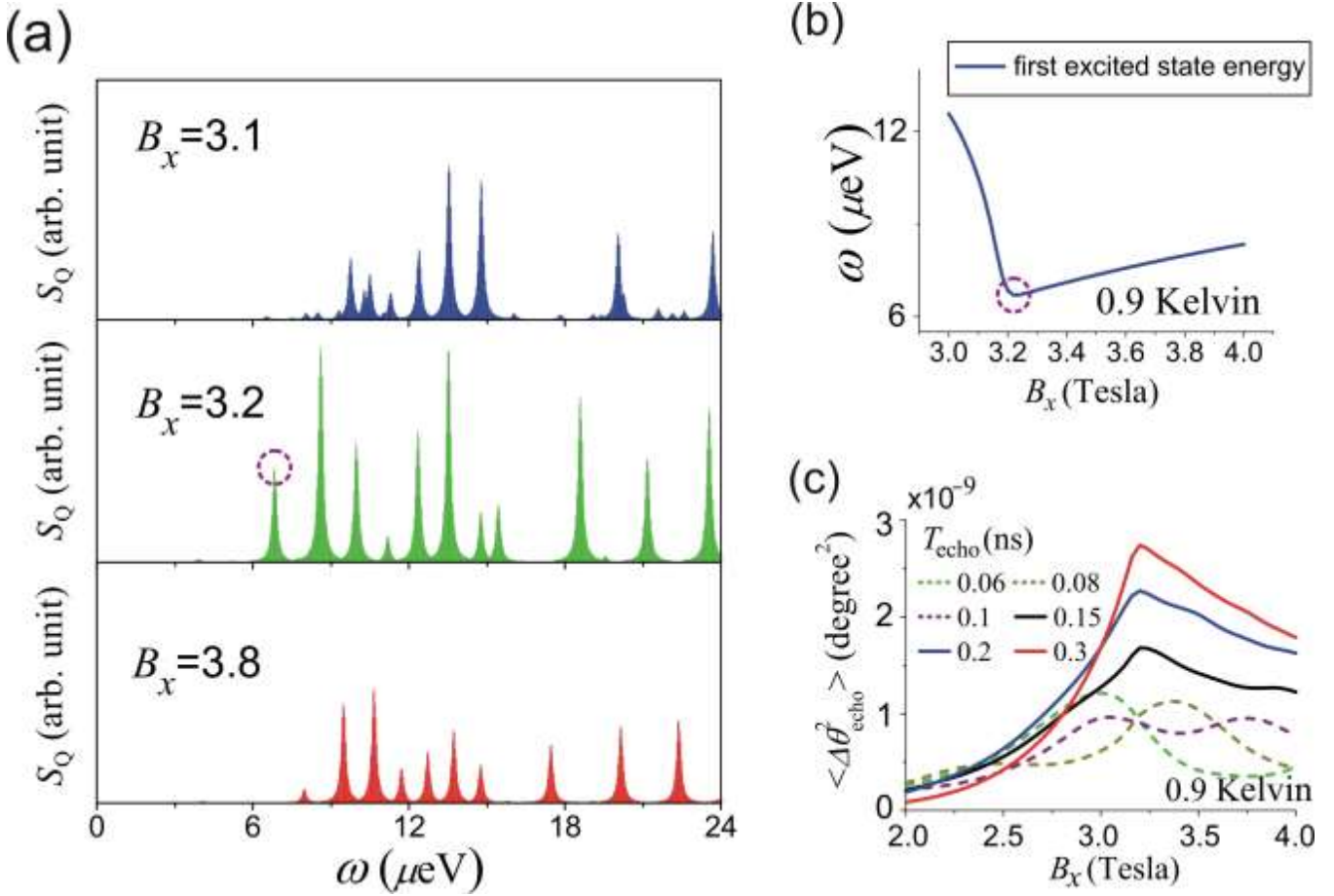

Figure 4 **Noise spectra and the effect of different interacting times.** (a) Noise spectra at $S_Q$ at temperature 0.9 Kelvin for various transverse magnetic fields. (b) The lowest excited state energy as a function of the transverse field at temperature 0.9 Kelvin. The purple circles in (a) and (b) mark the lowest energy excitation at transverse field around $B_x = 3.2$ Tesla. (c) $\langle \Delta\theta^2_{echo}\rangle$ as a function of transverse field at temperature 0.9 Kelvin for various interaction time $T_{echo}$ and delay time $\tau = 0$ ns. The sample thickness is set as $L_S = 0.5$ cm, and laser cross sectional area is set as $A = 10$ μm$^2$.

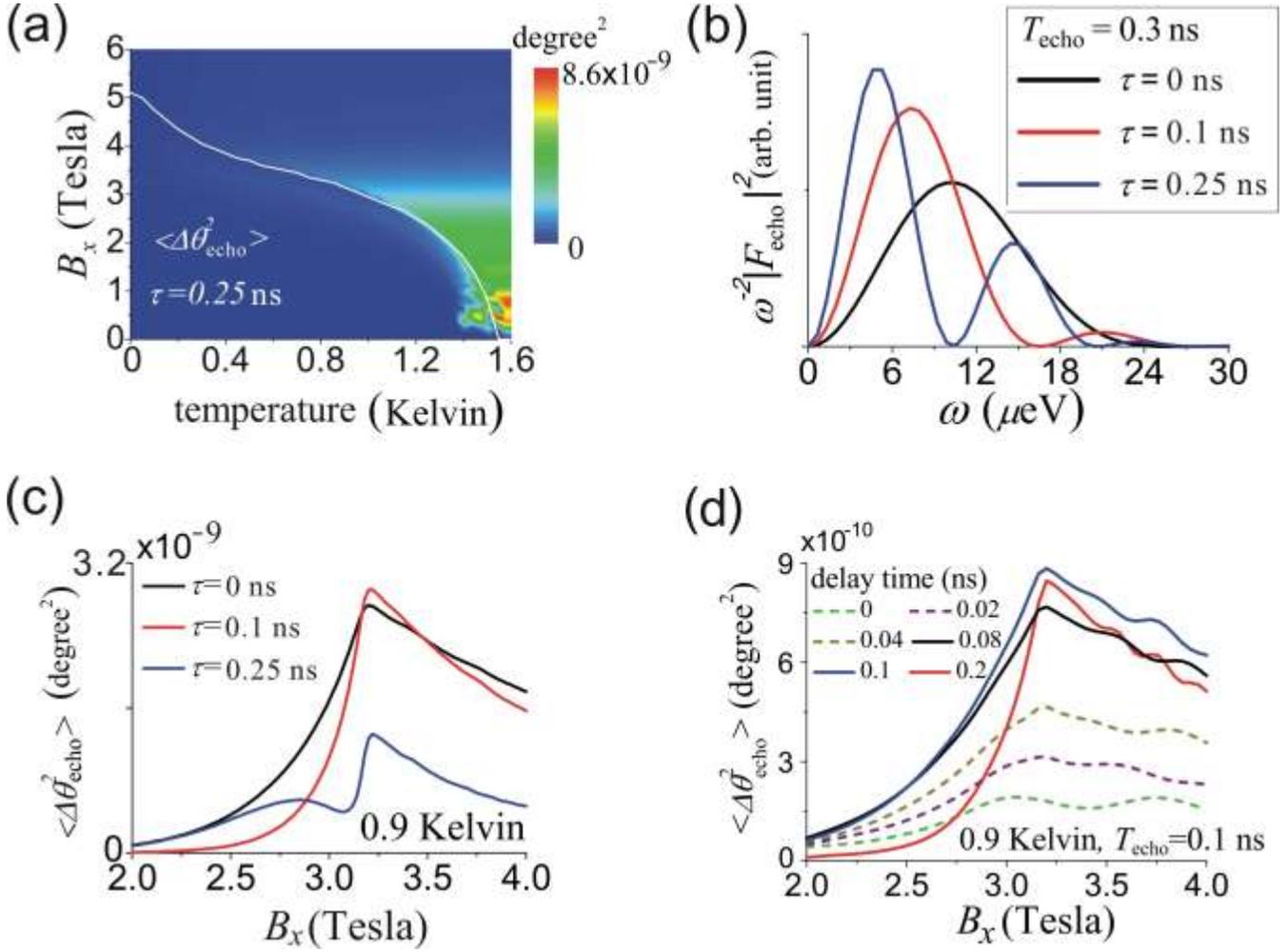

Figure 5 | **Effect of different delay times.** (a) $\langle \Delta\theta_{echo}^2 \rangle$ as a function of temperature and transverse field, for interaction time $T_{echo}$=0.3 ns and delay between two interaction intervals $\tau = 0.25$ ns. The white curve indicates the phase boundary. (b) The filter function $\omega^{-2}|F_{echo}|^2$ for $T_{echo}$=0.3 ns and various delay times $\tau$. (c) $\langle \Delta\theta_{echo}^2 \rangle$ as a function of transverse field at temperature 0.9 Kelvin for $T_{echo}$ =0.3 ns and various delay times $\tau$. (d) $\langle \Delta\theta_{echo}^2 \rangle$ as a function of transverse field at temperature 0.9 Kelvin for $T_{echo}$ =0.1 ns and various delay times $\tau$. The sample thickness is set as $L_S = 0.5$ cm in (a) & (c), and $L_S = 0.1$ cm in (d). The laser cross sectional area is set as $A = 10$ μm$^2$ in all the figures.